\documentclass[aps,prd,twocolumn,superscriptaddress,10pt,nofootinbib,floatfix]{revtex4-2}
\usepackage{amsfonts}
\usepackage{float}
\usepackage{eurosym}
\usepackage{amssymb}
\usepackage{textcomp}
\usepackage{graphicx}
\usepackage{epsfig}
\usepackage{subfigure}
\usepackage{amsmath}
\usepackage{slashed}
\usepackage{units}
\usepackage{hyperref}
\usepackage{array}
\usepackage{lmodern}
\usepackage{dcolumn}
\usepackage{bm}

\begin{document}

\title{Stellar Light Scattering as a Probe for a Braneworld-Induced Baryogenesis Scenario}

\author{Micha\"{e}l Sarrazin}
\email{michael.sarrazin@ac-besancon.fr}
\affiliation{Universit\'{e}
Marie et Louis Pasteur, CNRS, Institut UTINAM (UMR 6213), \'{E}quipe de Physique Th\'{e}orique, F-25000 Besan\c con, France}
\affiliation{Department of Physics, University of Namur, B-5000 Namur,
Belgium}

\begin{abstract}
A recent baryogenesis scenario [Phys. Rev. D 110, 023520 (2024)], rooted in a two-brane Universe model, proposed a solution to the matter-antimatter asymmetry through the dynamics of a new pseudo-scalar field. In the present paper, one investigates the phenomenological consequences of this proposal. One shows that the associated boson could persist as a relic from the early Universe, forming a subdominant component of dark matter. While its overall cosmological density is small ($\approx 0.2\%$), one demonstrates that a one-loop process facilitates an ultra-weak coupling to photons, leading to a distinctive scattering signature. One argues that this effect could produce a faint, glowing halo around massive, hot stars, characterized by a unique spectral decay. Detecting or constraining this elusive light with current and future instruments like the JWST would provide a powerful and direct observational test of the underlying braneworld dynamics and its connection to baryogenesis.
\end{abstract}

\maketitle

\section{Introduction}

\label{Intro}

Over the past decades, braneworld scenarios -- where our observable Universe 
is modeled as a 3-dimensional brane embedded in a higher-dimensional bulk -- have 
generated significant interest for their potential to address fundamental problems 
in cosmology and particle physics \cite%
{br1,br4,br6,dgp,br7}. In particular, when a hidden brane coexists 
near our visible brane in the bulk, new phenomena emerge with possible implications 
for the nature of dark matter and dark energy \cite{RevBr1, RevBr2, DMBrane1, DMBrane2,
Phant1, Phant2}, or even for the origin of the Big Bang itself through brane collisions \cite%
{Branerev,cob1,cob5,cob6,cob7,cob8,cob9,cob10,cob11}.

In a recent work \cite{baryo}, it was shown that such multi-brane models could naturally 
explain the observed matter-antimatter asymmetry, thereby addressing the baryogenesis 
conundrum \cite{baryoR,Bolt,rev1}. The mechanism invokes a new pseudo-scalar field 
associated with inter-brane dynamics, providing a novel pathway to generate the visible 
matter of our Universe. Testing the existence and properties of the corresponding boson 
would therefore provide a direct experimental handle on this specific baryogenesis mechanism 
and, more broadly, on the viability of braneworld cosmologies \cite{cob5}. This is the purpose of the present paper.

While the braneworld framework itself remains speculative, it offers an elegant 
theoretical structure for addressing longstanding questions. This work focuses on the 
observable consequences of the baryogenesis model proposed in \cite{baryo}. Our goal is to derive a clear, 
testable prediction that can be confronted with astrophysical observations, thereby 
providing a means to experimentally constrain this class of theories.

One begins by showing how the scalar boson, a key ingredient of the baryogenesis 
model, can be produced during the Big Bang and could persist today as a dark 
relic (Section III). Although this boson contributes only a minor 
fraction ($\approx 0.2\%$) to the total dark matter density, one demonstrates 
that it can interact with photons via a one-loop quantum process 
(Section IV). This interaction, however weak, enables a unique spectroscopic 
signature around massive stars, which could be detected with modern instruments \cite%
{axdm1,axdm2,axdm3,Navarro1997,Bertone2005,Gondolo2004,axdm4,axdm5,axdm6,axdm7}. In Section V, 
one details the characteristics of this signature and discuss the observational 
prospects for its detection. We argue that a dedicated search for this faint, 
scattered light could place meaningful constraints on the boson's properties 
and, by extension, on the braneworld scenario from which it originates.

\section{Theoretical framework}

\label{Theo}

In the present section, one just recalls the Lagrangian describing the
matter fields and the electromagnetic fields in a two-brane Universe, as
well as the scalar field of interest, and their couplings. For the full
derivation and theoretical demonstration of this Lagrangian, one refers the
reader to previous publications \cite{M4xZ2,cp1,cp2,baryo}. This framework builds 
on the equivalence between a two-brane universe embedded in 
a $(3+N,1)$-dimensional bulk and a noncommutative two-sheeted 
spacetime $M_4 \times Z_2$ \cite{M4xZ2,cp1,cp2}. Far from a phenomenological ansatz, 
this equivalence, rigorously proven in our prior works \cite{M4xZ2,cp1,cp2}, 
applies broadly to braneworld theories, from string models to domain walls frameworks.

To provide context, as shown in earlier work \cite{M4xZ2}, the scalar field 
represents the extra-dimensional component of the electromagnetic field in the bulk, 
dressed by a fermionic field following the Dvali-Gabadadze-Shifman 
mechanism \cite{dgp}. As a consequence, the model hypothesizes the existence of 
multiple effective scalar fields, each with distinct masses corresponding to 
specific quarks or leptons of a given flavor and generation in the Standard Model. 

The effective Lagrangian, presented below, implicitly includes a geometrical mixing term with a coupling 
constant $g = (m^2 / M_B) \exp(-m d) \sim m^2 / M_B$, where $M_B$ is the brane energy 
scale, $m$ is the mass of the fermion under consideration and $d$ is the interbrane distance. 
Both $M_B$ and $d$ are constrained by LHC searches for extra-dimensional 
signatures \cite{PDG}, such that $M_B \geq 15~\text{TeV}$ 
and $d \ll m$ \cite{cp1} -- i.e. $\exp(-m d) \sim 1$. This mixing enables interactions between 
visible and hidden branes, with potential experimental signatures such as neutron-hidden neutron 
oscillations explored in passing-through-wall experiments \cite{Exp0,Exp1,Exp2,Exp3,Exp4}.

Here, one considers scalar fields that could exist nowadays, i.e. stable
scalar fields, i.e. those which rely on stable quarks or leptons, or with
fermionic loops with low mass states \cite{PDG}. Consequently, only the
scalar field associated with the electron is relevant in the subsequent
discussion. This implies that such a scalar field can only manifest when
electrons can exist, which is from the beginning of the quark era and at
temperatures below $150$ GeV.

The Lagrangian of interest can be then written as: 
\begin{eqnarray}
\mathcal{L} &=&\mathcal{L}_{matter}+\mathcal{L}_{EM}+\mathcal{L}_{\varphi }
\label{Lag} \\
&&+\mathcal{L}_{EM-matter}+\mathcal{L}_{\varphi -matter} ,  \notag
\end{eqnarray}%
where: 
\begin{equation}
\mathcal{L}_{matter}=\overline{\psi }_{+}\left( i\gamma ^{\mu }\partial
_{\mu }-m\right) \psi _{+}+\overline{\psi }_{-}\left( i\gamma ^{\mu
}\partial _{\mu }-m\right) \psi _{-},  \label{Lagm}
\end{equation}%
is the sum of the Lagrangians of the Dirac fields of the electron in each
brane $(\pm )$, and: 
\begin{equation}
\mathcal{L}_{EM}=-\frac{1}{4}F^{+\mu \nu }F_{\mu \nu }^{+}-\frac{1}{4}%
F^{-\mu \nu }F_{\mu \nu }^{-},  \label{LagEM}
\end{equation}%
is the sum of the Lagrangians of the electromagnetic fields in each brane,
with $F_{\mu \nu }^{\pm }=\partial _{\mu }A_{\nu }^{\pm }-\partial _{\nu
}A_{\mu }^{\pm }$ ($A_{\mu }^{\pm }$ are the electromagnetic four-potentials
on each brane $(\pm ))$. The pseudo-scalar field $\varphi$ of mass $%
m_{\varphi }$ is common to both branes, and its Lagrangian is given by: 
\begin{equation}
\mathcal{L}_{\varphi }=\frac{1}{2}\left( \partial _{\mu }\varphi \right)
\left( \partial ^{\mu }\varphi \right) -\frac{1}{2}m_{\varphi }^{2}\varphi
^{2}-V(\varphi),  \label{Lagscal}
\end{equation}%
with: 
\begin{equation}
V(\varphi )=\frac{1}{2}em_{\varphi }\varphi ^{3}+\frac{1}{8}e^{2}\varphi
^{4},  \label{self}
\end{equation}%
the self-coupling terms of $\varphi $ and where $e$ is the electromagnetic
coupling constant. Also, one shows that \cite{cp1,cp2,baryo}: 
\begin{equation}
m_{\varphi } = 2g = \frac{2m^{2}}{M_{B}},  \label{mass}
\end{equation}%
with $M_{B}^{-1}$ the brane effective thickness \cite{cp1,cp2}. Finally,
some couplings occur:%
\begin{equation}
\mathcal{L}_{EM-matter}=e\overline{\psi }_{+}\gamma ^{\mu }\psi _{+}A_{\mu
}^{+}+e\overline{\psi }_{-}\gamma ^{\mu }\psi _{-}A_{\mu }^{-},
\label{LagEMmat}
\end{equation}%
is the sum of the usual interaction terms between the Dirac fields --
endowed with the charge $-e$ -- and the electromagnetic fields for each
brane, and:%
\begin{equation}
\mathcal{L}_{\varphi -matter}=\frac{1}{2}e\mathcal{W}_{+}^{\dag }\overline{%
\psi }_{+}\gamma ^{5}\mathcal{W}_{-}\psi _{-}\varphi -\frac{1}{2}e\mathcal{W}%
_{-}^{\dag }\overline{\psi }_{-}\gamma ^{5}\mathcal{W}_{+}\psi _{+}\varphi,
\label{LagInt}
\end{equation}%
is the coupling between the pseudo-scalar field and the Dirac fields of each
brane \cite{baryo} with $\mathcal{W}$ the Wilson line \cite{WL1,WL2}:%
\begin{equation}
\mathcal{W}_{\pm }=\mathcal{P}\left\{ \exp \left( -ie\int_{C}A_{\mu }^{\pm
}dx^{\mu }\right) \right\}.  \label{wilson}
\end{equation}

The consequences of the Lagrangian (\ref{Lag}) are explored in the next
sections.

\section{Massive scalar bosons as a relic of the Big Bang}

\label{Relic}

\begin{figure*}[ht!]
\centerline{\ \includegraphics[width=17cm]{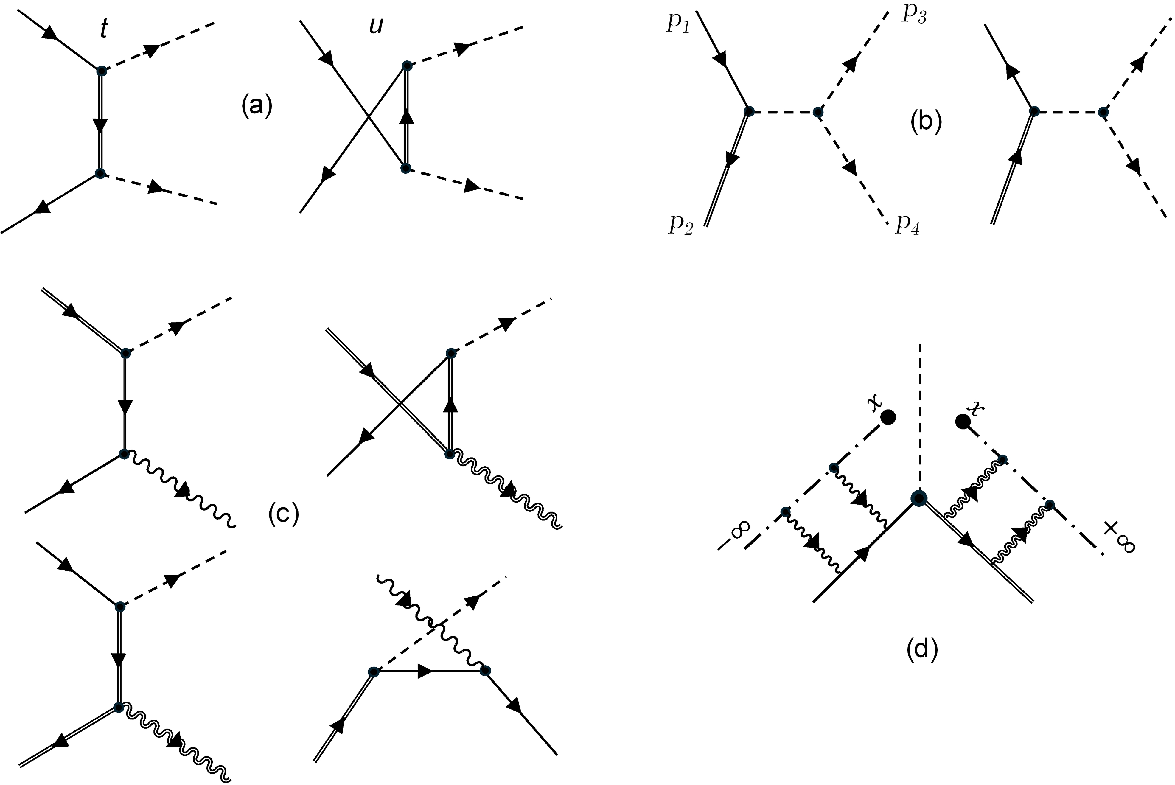}} \vspace*{8mm}
\caption{Feynman diagrams of interest. Simple wavy line: photon ; double
wavy line: hidden photon ; simple straight line: fermion ; double straight
line: hidden fermion ; dashed line: scalar boson ; dashed dotted line:
Wilson line. (a): Fermion-antifermion annihilation into a boson pair. (b):
Fermion-hidden antifermion (or hidden fermion-antifermion) annihilation into
a boson pair. (c): Various fermion annihilation processes involving a boson,
a photon (or hidden photon), or a fermion. (d): Order-two correction from
the Wilson lines to the vertex of the fermion-boson interaction.}
\label{fig:f1}
\end{figure*}

In the present section, one considers the production of scalar bosons during
the first moments of the Big Bang and how much they might still exist today
as relics. Assuming the Lagrangian (\ref{Lag}) as well as the conservation
of momentum, spin, charge and flavors, the scalar boson is stable. It is
also important to note that only charged particles -- such as electrons or
positrons here -- can interact with the scalar boson, while neutral
particles like neutrinos remain unaffected by its presence. Then, the
typical lowest-order Feynman diagrams for fermion-antifermion annihilation
are shown in Fig. 1 (a), (b) and (c) and implies two vertices. Other
annihilation channels with more than two vertices are not considered here. A
careful analysis of diagrams (c) shows that they violate the Ward
identities. This is due to the Wilson lines mentioned above, and which
cannot be avoided even at the lowest order \cite{WL1,WL2}. For instance, the
vertex for the boson-fermion interaction must be corrected \cite{WL1,WL2} as
shown in diagram (d) of the Fig.1. Here, the vertex correction at the second
order is shown but all other orders should be considered. As a consequence,
the correct description of the mechanisms suggested by diagrams (c) involves
more than two vertices and are not considered in the following. Now, a
comprehensive analysis of diagrams (a) indicates a null contribution due to
spin conservation in relation to the pseudo-scalar properties of the boson.
This can be verified from a full computation of the amplitude $\mathcal{M}$
implying diagrams (a) \cite{scalpot}. By contrast, diagrams (b) easily leads
to:

\begin{equation}
\left\vert \mathcal{M}_{\lambda \lambda ^{\prime }}\right\vert ^{2}=\frac{%
\pi ^{2}\alpha _{\text{em}}^{2}m_{\varphi }^{2}}{\left( s-m_{\varphi
}^{2}\right) ^{2}}s\delta _{\lambda \lambda ^{\prime }},  \label{Amp2}
\end{equation}%
with $\alpha _{\text{em}}=e^{2}/(4\pi )$ the fine-structure constant, $s$
the usual Mandelstam variable, and where $\lambda $ and $\lambda ^{\prime }$
denote the polarization states of each fermion. Then the average squared
amplitude follows:%
\begin{eqnarray}
\overline{\left\vert \mathcal{M}\right\vert ^{2}}\left(
e_{h}^{+}e^{-}\rightarrow 2\varphi \right) &=&\overline{\left\vert \mathcal{M%
}\right\vert ^{2}}\left( e^{+}e_{h}^{-}\rightarrow 2\varphi \right)
\label{Mm2b} \\
&=&\frac{\pi ^{2}\alpha _{\text{em}}^{2}m_{\varphi }^{2}}{2\left(
s-m_{\varphi }^{2}\right) ^{2}}s,  \notag
\end{eqnarray}%
where one has considered unpolarized electrons and positrons. One can note
that $\overline{\left\vert \mathcal{M}\right\vert ^{2}}\left( 2\varphi
\rightarrow e_{h}^{+}e^{-}\right) =\overline{\left\vert \mathcal{M}%
\right\vert ^{2}}\left( 2\varphi \rightarrow e^{+}e_{h}^{-}\right) =4%
\overline{\left\vert \mathcal{M}\right\vert ^{2}}\left(
e_{h}^{+}e^{-}\rightarrow 2\varphi \right) =4\overline{\left\vert \mathcal{M}%
\right\vert ^{2}}\left( e^{+}e_{h}^{-}\rightarrow 2\varphi \right) $. Let us
now consider the cross-section $\sigma \left( s\right) =\sigma \left(
e_{h}^{+}e^{-}\rightarrow 2\varphi \right) =\sigma \left(
e^{+}e_{h}^{-}\rightarrow 2\varphi \right) $, one gets:%
\begin{equation}
\sigma \left( s\right) =\frac{\pi ^{2}\alpha _{\text{em}}^{2}m_{\varphi }^{2}%
}{32\pi }\frac{1}{\left( s-m_{\varphi }^{2}\right) ^{2}}\frac{\sqrt{%
s-4m_{\varphi }^{2}}}{\sqrt{s-4m^{2}}}.  \label{sii}
\end{equation}%
Considering a fermion gas with a thermal velocity distribution, the relevant
parameter is the annihilation rate $\left\langle \sigma v\right\rangle $
(with the velocity $v$) given by the well-known relation \cite{sg}:

\begin{equation}
\left\langle \sigma v\right\rangle =\frac{1}{8Tm^{4}K_{2}^{2}(\frac{m}{T})}%
\int_{4m^{2}}^{\infty }ds\sqrt{s}(s-4m^{2})K_{1}(\frac{\sqrt{s}}{T})\sigma
\left( s\right) .  \label{rate}
\end{equation}%
Then, after some simplifications and rearrangements, one gets:

\begin{equation}
\left\langle \sigma v\right\rangle =\left\langle \sigma v\right\rangle
_{0}f(\varkappa ),  \label{ratef}
\end{equation}%
with:%
\begin{equation}
\left\langle \sigma v\right\rangle _{0}=\frac{\pi \alpha _{\text{em}}^{2}}{64%
}\frac{1}{M_{B}^{2}},  \label{rate0}
\end{equation}%
and where:%
\begin{equation}
f(\varkappa )=\frac{2}{K_{2}^{2}(\varkappa /2)}\int_{\varkappa }^{\infty
}K_{1}(y)\sqrt{1-\frac{\varkappa ^{2}}{y^{2}}}dy,  \label{func}
\end{equation}%
with $\varkappa =2m/T$ with $\lim_{\varkappa \rightarrow +\infty
}f(\varkappa )=1$ and $\lim_{\varkappa \rightarrow 0}f(\varkappa )=0$. From
previous work, $M_{P}/100<M_{B}<M_{P}$ to match with experimental
constraints \cite{cp1}. However, so as not to limit ourselves unnecessarily,
two cases can be considered, first when $M_{B}=M_{P}$, and next -- according
to the limits from the Large Hadron Collider on extradimensions \cite{PDG}
-- when $M_{B}=15$ TeV. Anyway, this leads to an extremely weak value for $%
\left\langle \sigma v\right\rangle _{0}$ and thus $\left\langle \sigma
v\right\rangle $.

Usually, the Boltzmann transport equation \cite{Bolt2, PU} leads to the
Lee-Weinberg equations \cite{LW} that govern the density of relic particles
in the expanding Universe. The density of scalar bosons $n_{\varphi }$ thus
would obey to \cite{Bolt2, PU}:%
\begin{equation}
\partial _{t}n_{\varphi }+3Hn_{\varphi }=-\left\langle \sigma \left(
2\varphi \rightarrow f\overline{f}\right) v\right\rangle \left( n_{\varphi
}^{2}-n_{\varphi ,eq}^{2}\right),  \label{BTE}
\end{equation}%
with $H$ the Hubble parameter, $\sigma $ the scalar boson annihilation
cross-section, $v$ the relative velocity between particles, and $%
\left\langle \cdots \right\rangle $ the thermal average at temperature $T$.
Quantity $n_{\varphi ,eq}$ is at the thermal equilibrium and is described by
the Bose-Einstein statistics. Then, in the well-known freeze-out mechanism,
when: 
\begin{equation}
n_{\varphi }\left\langle \sigma v\right\rangle <H,  \label{cond}
\end{equation}%
there is no more thermal equilibrium between the scalar bosons and the
fermions. Here, from Eq. (\ref{func}), it can be shown that (\ref{cond}) is
always verified at any time, such that the scalar boson population can never
be at the equilibrium. Rather, a freeze-in scenario \cite{FI1,FI2,FI3} must
be considered where $n_{\varphi }=0$ when the electron-positron population
appear in the primordial plasma, and where the scalar bosons only appear due
to fermion annihilation. Then, $n_{\varphi }$ follows:

\begin{eqnarray}
\partial _{t}n_{\varphi }+3Hn_{\varphi } &=&\left\langle \sigma
v\right\rangle n_{e_{h}^{+},eq}n_{e^{-},eq}+\left\langle \sigma
v\right\rangle n_{e_{h}^{-},eq}n_{e^{+},eq}  \notag \\
&=&2\left\langle \sigma v\right\rangle n_{f,eq}^{2},  \label{FII}
\end{eqnarray}%
where $f_{f,eq}$ is described by the Fermi-Dirac statistics and such that $%
n_{e_{h}^{+},eq}=n_{e^{-},eq}=n_{e_{h}^{-},eq}=n_{e^{+},eq}=f_{f,eq}$. One
sets the comoving particle density $Y_{\varphi }=n_{\varphi }/s$, where $s$
is the entropy density such that:

\begin{equation}
s=\frac{16\pi ^{2}}{45}m^{3}q_{\ast }\varkappa ^{-3},  \label{s}
\end{equation}%
with $q_{\ast }$ the effective number of degrees of freedom defined for the
entropy density. Then, Eq. (\ref{FII}) can be rewritten as:%
\begin{equation}
\frac{dY_{\varphi }}{d\varkappa }=2\left\langle \sigma v\right\rangle \frac{%
s\eta }{Hx}Y_{f,eq}^{2},  \label{FIEq}
\end{equation}%
where one used $dx/dt=Hx/\eta $, with: \cite{PU}%
\begin{equation}
Y_{f,eq}=\frac{45}{16\pi ^{4}q_{\ast }}\varkappa ^{3}\int_{1}^{\infty }\frac{%
u\sqrt{u^{2}-1}}{1+\exp \left( \frac{\varkappa }{2}u\right) }du,  \label{Yeq}
\end{equation}%
and: 
\begin{equation}
\eta =1-\frac{\varkappa }{3q_{\ast }}\frac{dq_{\ast }}{d\varkappa }.
\label{ad}
\end{equation}%
While $\eta $ is often close to $1$ during most of the radiation era, it is
not the case, for instance, shortly after the QGPHG transition as pions and
muons annihilate between $160$ MeV and $100$ MeV leading then to a fast
change of $q_{\ast }$ against $x$. In the same way, since the period of
interest is radiatively-dominated, the Hubble parameter is defined through 
\cite{Bolt2, PU}:%
\begin{equation}
H=\frac{8\pi \sqrt{\pi }}{3\sqrt{5}}\frac{m^{2}}{M_{P}}g_{\ast
}^{1/2}\varkappa ^{-2},  \label{Hubble}
\end{equation}%
with $g_{\ast }$ the effective number of degrees of freedom defined for the
energy density, and where $M_{P}$ is the Planck mass. Both functions $%
g_{\ast }$ and $q_{\ast }$ can be fitted from exact computations \cite{DF}
and one can set $g_{\ast }=q_{\ast }$ \cite{Bolt2, PU,DF}. Then, from Eq. (%
\ref{FIEq}) one easily gets: 
\begin{equation}
Y_{\varphi }=\int_{\varkappa _{0}}^{\varkappa }2\left\langle \sigma
v\right\rangle \frac{s\eta }{Hx}Y_{f,eq}^{2}d\varkappa ^{\prime },  \label{Y}
\end{equation}%
which can be simply computed numerically as shown in Fig. 2 where: 
\begin{equation}
Y_{\varphi ,0}=\frac{135\sqrt{5}\alpha _{\text{em}}^{2}}{2048\pi ^{6}\sqrt{%
\pi }}\frac{mM_{P}}{M_{B}^{2}},  \label{Y0}
\end{equation}%
and with $\varkappa _{0}=6.8\times 10^{-6}$, i.e. $T_{0}=150$ GeV.

\begin{figure}[th]
\centerline{\ \includegraphics[width=8.5cm]{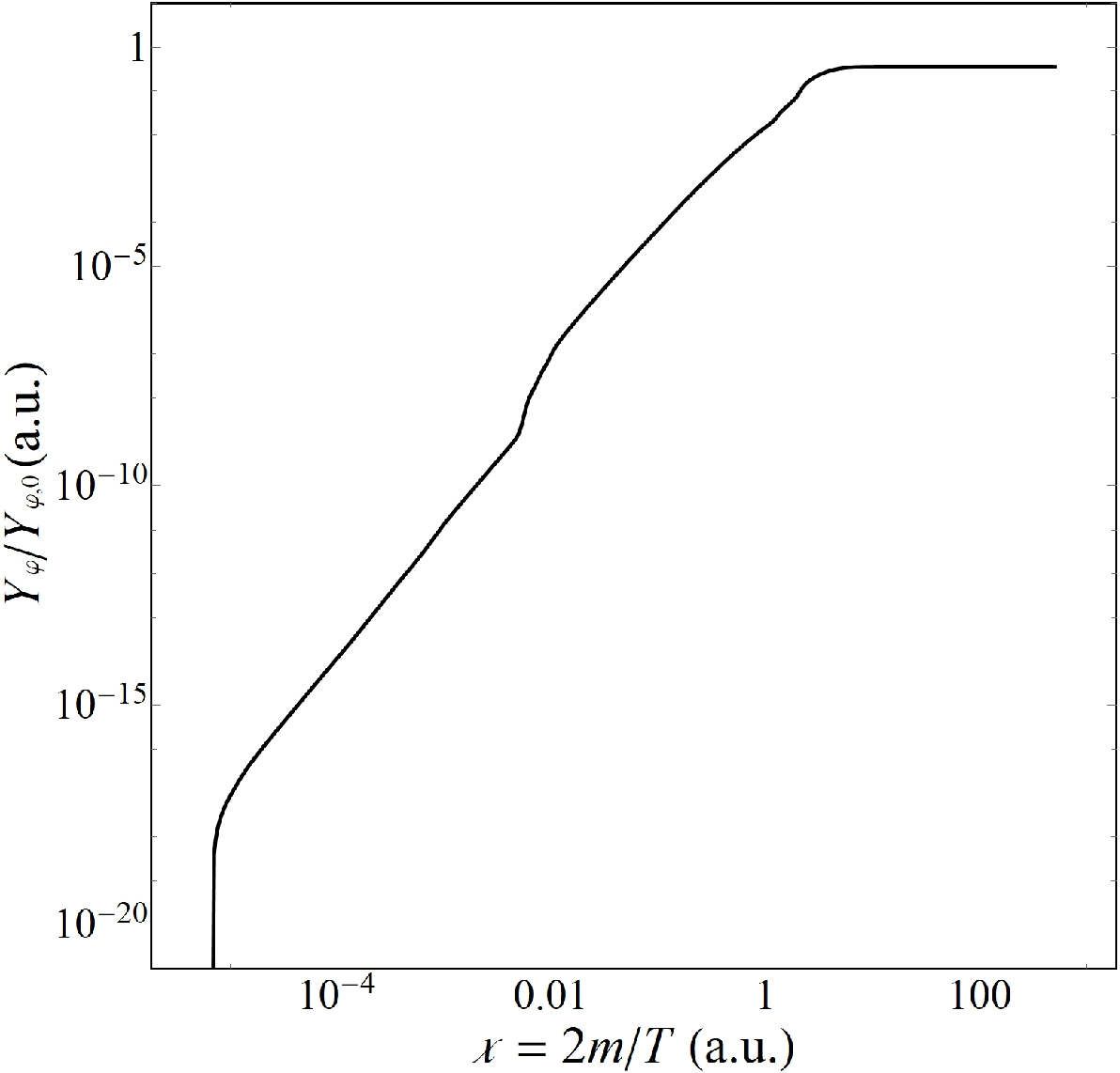}} \vspace*{8mm}
\caption{Scalar boson relative comoving abundance against $\varkappa =2m/T$,
showing a typical freeze-in behavior.}
\label{fig:f2}
\end{figure}

\begin{figure*}[t]
\centerline{\ \includegraphics[width=18cm]{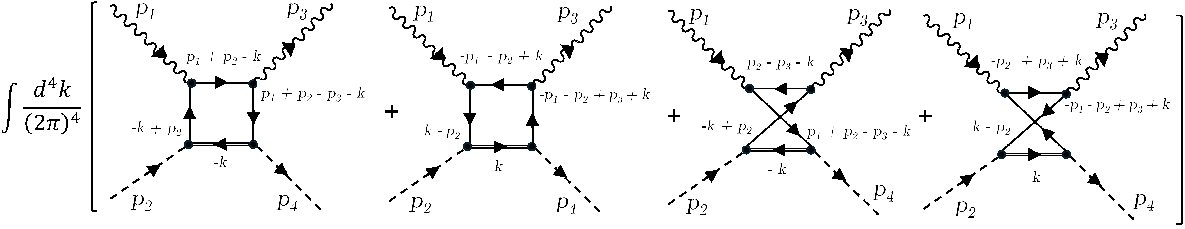}} \vspace*{8mm}
\caption{Feynman diagrams contributing to the amplitude of the pseudo-scalar
boson-photon scattering. Simple wavy line: photon ; simple straight line:
fermion ; double straight line: hidden fermion ; dashed line: scalar boson.
The momentum of each incoming and outgoing particle is mentioned ($%
p_{1},p_{2},p_{3}$ and $p_{4}$), as well as the momentum flow $k$ along the
loop.}
\label{fig:f3}
\end{figure*}

From computations shown in Fig. 2, one gets:%
\begin{eqnarray}
Y_{\varphi ,\infty } &=&0.359Y_{\varphi ,0}  \label{Yinf} \\
&\sim &1.65\times 10^{-9}\frac{mM_{P}}{M_{B}^{2}}.  \notag
\end{eqnarray}%
One notes that $Y_{\varphi }$ has fully converged towards $Y_{\varphi
,\infty }$ for $T\geq T_{D}=m/4.$ Also, about 76 \% of the whole scalar
boson population is produced during the period such that $T\in \left[ m/4%
\text{ , }m\right] $. As $n_{\varphi ,today}=Y_{\varphi ,\infty }s_{today}$,
one gets:%
\begin{equation}
n_{\varphi ,today}=Y_{\varphi ,\infty }s_{today}=2.44\frac{mM_{P}}{M_{B}^{2}}%
\text{ m}^{-3}.  \label{ntod}
\end{equation}%
As an illustration, two cases can be considered, first when $M_{B}=M_{P}$,
and then:%
\begin{equation}
n_{\varphi ,today}\sim 1.0\times 10^{-22}\text{ m}^{-3}=3.4\times 10^{11}%
\text{ au}^{-3},  \label{cas1}
\end{equation}%
and next -- according to the limits from the Large Hadron Collider on
extradimensions -- when $M_{B}=15$ TeV, and then: 
\begin{equation}
n_{\varphi ,today}\sim 6.8\times 10^{7}\text{ m}^{-3}.  \label{cas2}
\end{equation}%
Also, one notes that the temperature of the relic today is roughly given by 
\cite{Bolt2, PU,DF}:%
\begin{equation}
\frac{T_{\varphi ,today}}{T_{CMB}}=\left( \frac{q_{\varphi }(T_{D})}{%
q_{\varphi }(T_{CMB})}\frac{q_{\gamma }(T_{CMB})}{q_{\gamma }(T_{D})}\right)
^{1/3},  \label{rel}
\end{equation}%
such that:%
\begin{equation}
T_{\varphi ,today}\approx 2\text{ }T_{CMB}\approx 5\times 10^{-4}\text{ eV.}
\label{temp}
\end{equation}%
From Eq. (\ref{mass}), the scalar mass $m_{\varphi }\approx 4.3\times
10^{-17}$ eV at $M_{B}=M_{P}$, and $m_{\varphi }\approx 35$ meV at $M_{B}=15$
TeV.\ In the first case, the scalar bosons are relativistic, and constitute
a hot component to dark matter, whereas, in the second case, they are cold
dark matter. One notes that the limit for the brane energy scale $M_{B}$
below which the scalar boson relic is cold is about $10^{4}$ TeV. But in
both previous cases, the scalar boson under consideration can only be a weak
component of the dark matter \cite{DM1,DM2,DM3}. Indeed, in the second case,
leading to the highest value of energy density for the scalar boson, one
gets $\rho _{\varphi }\sim m_{\varphi }n_{\varphi ,today}\approx 2.4\times
10^{-9}$ GeV cm$^{-3}$ i.e. about $0.2$ \% of the whole dark matter density $%
\rho _{DM}=1.26\times 10^{-6}$ GeV cm$^{-3}$ \cite{DM2}. Anyway, although
beyond the present topic, thanks to gravitational capture, this density
could be higher by many order of magnitude in the Milky Way or in our Solar
System. For instance, in Solar System the dark matter density is expected to
be of the order of $0.4$ GeV cm$^{-3}$ \cite{DM}, i.e. five order of
magnitude greater than the average dark matter density of the whole
Universe. In proportion, one could then expect for local values of $%
n_{\varphi ,today}$ such that: 
\begin{equation}
n_{\varphi ,local}\sim 2.2\times 10^{13}\text{ m}^{-3}.  \label{loc}
\end{equation}%
This local density, amplified by gravitational capture in the galactic 
halo \cite{Bertone2005}, is approximately five orders of magnitude higher 
than the cosmological average, enabling the detection of photon-boson 
interactions despite the boson’s minor contribution ($0.2$ \% of $\rho_{DM}$) 
to the total dark matter density.
Anyway, the scalar boson contribution as a dark matter component remains
very weak and is likely negligible from a cosmological point of view.
Nevertheless, as shown hereafter, as it is supposed to interact with light
through one-loop quantum processes, the boson relic could be detected to
constrain the scalar field. While the scalar boson model at $M_{B}=M_{P}$
leaves little hope of detection due to the extremely weak density, in the
limit $M_{B}<10^{4}$ TeV the boson-photon scattering could lead to
astrophysical signatures.

\section{Boson-Photon scattering}

\label{Scatt}

\begin{figure}[th]
\centerline{\ \includegraphics[width=8.5cm]{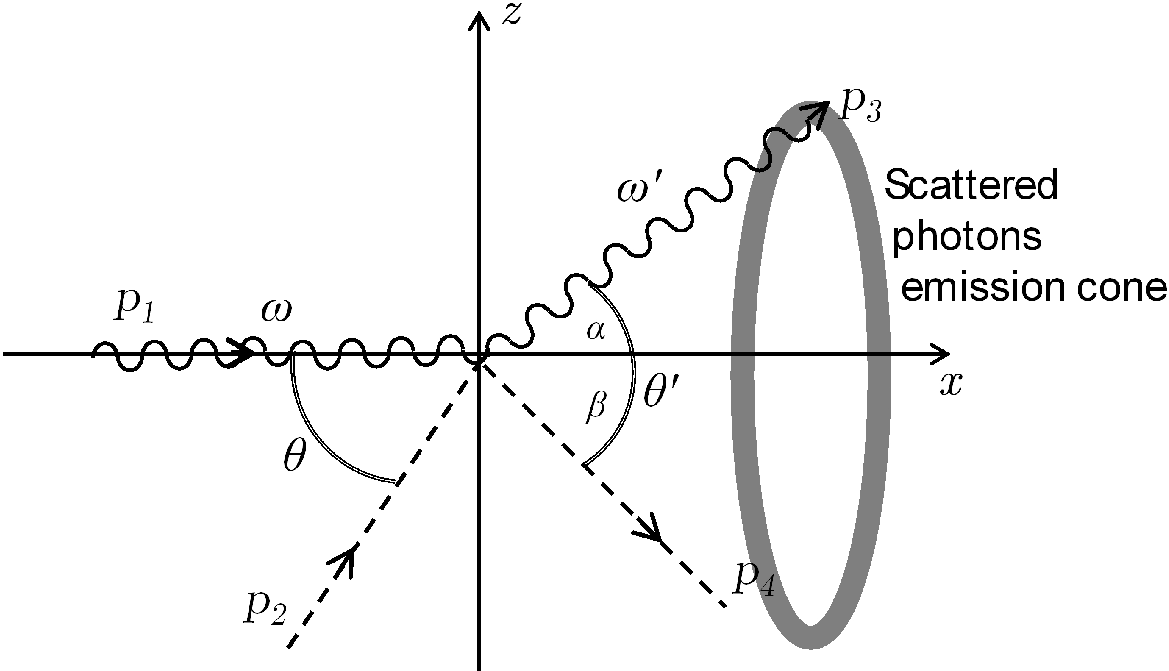}} \vspace*{8mm}
\caption{Pseudo-scalar boson--photon scattering and frame used to describe
the scattering. Simple wavy line: photon ; dashed line: scalar boson. The
momentum of each incoming and outgoing particle is mentioned ($%
p_{1},p_{2},p_{3}$ and $p_{4}$) with $\left\vert \mathbf{p}_{1}\right\vert =%
\protect\omega $ and $\left\vert \mathbf{p}_{3}\right\vert =\protect\omega %
^{\prime }$.}
\label{fig:f4}
\end{figure}

As explained in the previous section, the present scalar boson is not
abundant enough to be a good dark matter candidate alone. Nevertheless, one
can expect to detect it thanks to light diffusion through it. Following the
Lagrangian (\ref{Lag}), there is no direct coupling between the
electromagnetic field and the scalar boson. Nevertheless, it is possible to
build a photon-boson coupling at the one-loop level as shown in Fig. 3 (see
also Appendix A). Such a process is a Compton-like diffusion (see Fig. 4).
In the used frame (see Fig. 4), the four four-momenta are given by: 
\begin{eqnarray}
p_{1} &=&\left( 
\begin{array}{c}
\omega \\ 
\omega \\ 
0 \\ 
0%
\end{array}%
\right) \text{ ; }p_{2}=\left( 
\begin{array}{c}
E \\ 
p\cos \theta \\ 
0 \\ 
p\sin \theta%
\end{array}%
\right) \text{; }  \label{mom1} \\
p_{3} &=&\left( 
\begin{array}{c}
\omega ^{\prime } \\ 
\omega ^{\prime }\cos \alpha \\ 
0 \\ 
\omega ^{\prime }\sin \alpha%
\end{array}%
\right) \text{ ; }p_{4}=\left( 
\begin{array}{c}
E^{\prime } \\ 
p^{\prime }\cos \beta \\ 
0 \\ 
-p^{\prime }\sin \beta%
\end{array}%
\right) ,  \label{mom2}
\end{eqnarray}%
with: $p_{1}^{2}=p_{3}^{2}=0$ and $p_{2}^{2}=p_{4}^{2}=m_{\varphi }^{2}$.\
Using the momenta conservation, leading to: $p_{4}^{2}=\left(
p_{1}+p_{2}-p_{3}\right) ^{2}$, as well as the mass-shell conditions, one
easily gets for the pulsation $\omega ^{\prime }$ of the scattered light:%
\begin{equation}
\omega ^{\prime }=\frac{\omega \left( E-\sqrt{E^{2}-m_{\varphi }^{2}}\cos
\theta \right) }{\omega \left( 1-\cos \alpha \right) +\left( E-\sqrt{%
E^{2}-m_{\varphi }^{2}}\cos \left( \alpha -\theta \right) \right) }.
\label{wp}
\end{equation}%
The differential cross-section of the process under consideration is given
by:

\begin{equation}
\frac{d\sigma }{d\Omega }=\frac{\left\vert \mathbf{p}_{3}\right\vert }{64\pi
^{2}\left\vert p_{1}p_{2}\right\vert \left( \omega +E\right) }\overline{%
\left\vert \mathcal{M}\right\vert ^{2}},  \label{csc}
\end{equation}%
with $d\Omega =2\pi \sin \alpha \,d\alpha $ and where the average squared
amplitude over the initial photon polarization, and sum over the final
photon polarization is given by (see Appendix A): 
\begin{equation}
\overline{\left\vert \mathcal{M}\right\vert ^{2}}=\frac{20}{9}\frac{\alpha _{%
\text{em}}^{4}}{m^{4}}(p_{1}p_{3})^{2},  \label{Mc}
\end{equation}%
leading to the expression:%
\begin{equation}
\frac{d\sigma }{d\Omega }=\frac{5\alpha _{\text{em}}^{4}}{144\pi ^{2}m^{4}}%
\frac{\omega ^{\prime 3}\left( 1-\cos \alpha \right) ^{2}}{E-\sqrt{%
E^{2}-m_{\varphi }^{2}}\cos \theta }.  \label{ds}
\end{equation}%
One explores the consequences of this Compton-like process and of its
cross-section in the next section.

\section{Spectral signatures around stars}

\label{star}

To constrain the proposed model, a challenging yet significant objective would
be to detect a luminous halo of scalar bosons exhibiting a specific 
spectrum, around a star, as described in the following. 

\begin{figure}[th]
\centerline{\ \includegraphics[width=8.5cm]{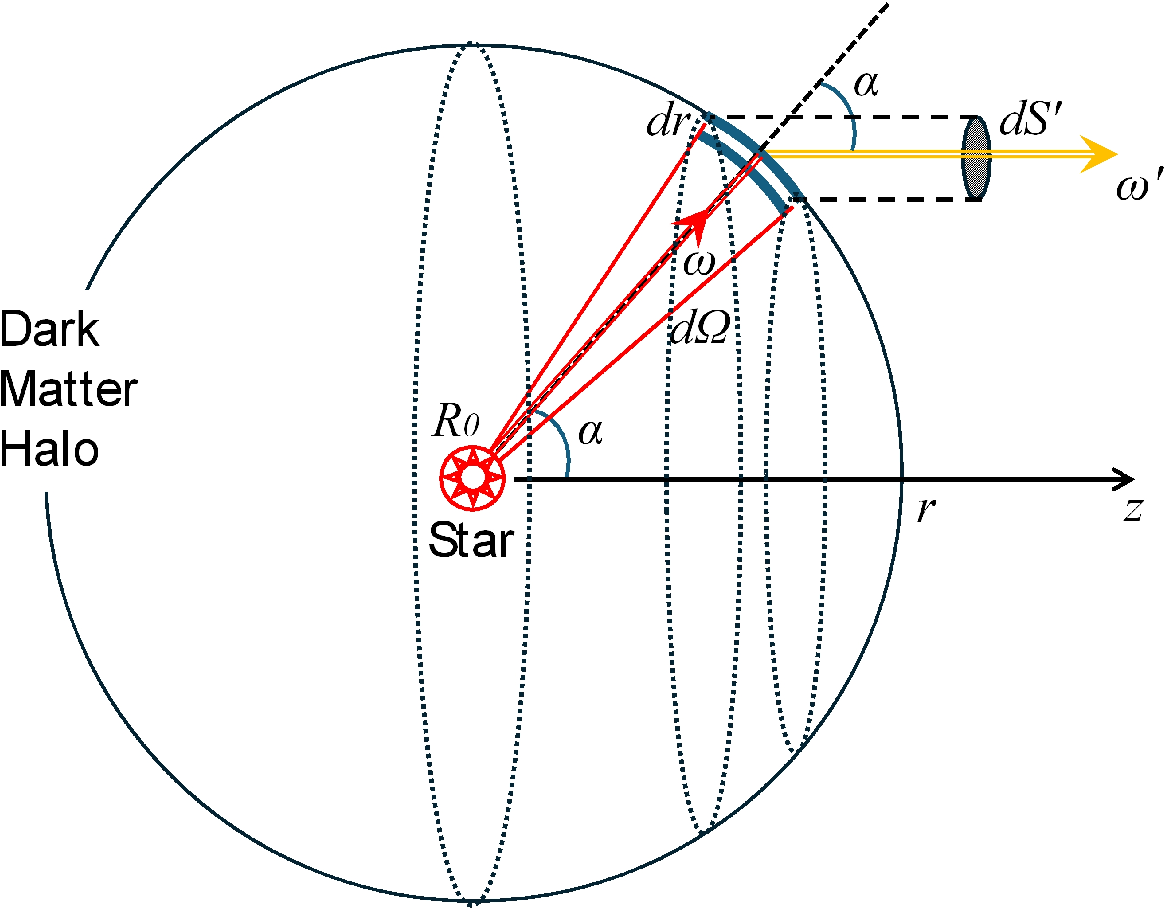}} \vspace*{8mm}
\caption{Sketch of the light diffusion by the scalar bosons as a dark matter
component around a star. The light from the star at pulsation $\protect%
\omega $ is scattered into photons at pulsation $\protect\omega ^{\prime }$. 
$Oz$ axis goes from the star center to the observer. $R_{0}$ is the star's
radius. $r$ is the radius of a shell of the scalar dark matter halo and
varies between $R_{0}$ and $+\infty $.}
\label{fig:f5}
\end{figure}

One considers a star modeled as a spherical source with a spectral radiance
given by $B(\omega )(R_{0}/r)^{2}$. For the halo of scalar bosons, one adopts
a Navarro-Frenk-White (NFW) density profile \cite{Navarro1997}, expressed
as: 
\begin{equation}
n(r)=\frac{n_{s}}{\frac{r}{r_{s}}(1+\frac{r}{r_{s}})^{2}}  \label{density}
\end{equation}%
where $n_{s}=4n_{\varphi ,local}$ is calibrated to match the local density $%
n_{\varphi ,local}\sim 2.2\times 10^{13}$ m$^{-3}$ due to gravitational
capture and $r_{s}\sim R_{0}$ the radius of the star. This profile,
motivated by cosmological simulations and studies of gravitational capture 
\cite{Bertone2005, Gondolo2004}, avoid the simplifying assumption of a
uniform distribution, concentrating the dark matter near the star where the
photon flux is maximal. The photon flux $\Phi (\omega )$ is given by:%
\begin{equation}
\Phi \left( \omega \right) =\left( \frac{R_{0}}{r}\right) ^{2}\frac{B(\omega
)}{\omega },  \label{PHI}
\end{equation}%
with the black body spectral radiance:%
\begin{equation}
B\left( \omega \right) =\frac{1}{4\pi ^{3}}\frac{\omega ^{3}}{\exp \left(
\omega /T\right) -1}.  \label{spec}
\end{equation}

As a toy model, and in a first approximation, one considers a dark matter
halo with a uniform distribution around the star. Then, the number of
photons $\delta N_{\omega ^{\prime }}$ at fixed frequency $\omega ^{\prime }$
produced per unit of time through the surface $dS^{\prime }$ in direction to
the observer and by a volume $r^{2}drd\Omega $ of scalar bosons with density 
$n_{\varphi }$ (see Fig. 5) is given by:%
\begin{equation}
\delta N_{\omega ^{\prime }}=\left( \frac{d\sigma }{d\Omega ^{\prime }}%
\right) \left( \frac{d\Omega ^{\prime }}{dS^{\prime }}\right) dS^{\prime
}\times \Phi \left( \omega \right) d\omega \times n_{\varphi }r^{2}dr\left( 
\frac{d\Omega }{d\alpha }\right) d\alpha ,  \label{dN}
\end{equation}%
with $dS^{\prime }=r^{2}d\Omega ^{\prime }\cos \alpha $. Then, the full
spectral radiosity $J$ is given by:%
\begin{equation}
J=\int_{V_{\omega ^{\prime }}}\omega ^{\prime }\frac{\delta N_{\omega
^{\prime }}}{dS^{\prime }d\omega ^{\prime }},  \label{Jb}
\end{equation}%
where $V_{\omega ^{\prime }}$ is the volume of space for which the range of
values of $\alpha $ defining the solid angle $\Omega $ are constrained by
the condition $\omega ^{\prime }=\omega ^{\prime }(\omega ,\alpha )$ given
by Eq. (\ref{wp}). For a fixed value $\omega ^{\prime }$, that means the
values of $\omega $ -- leading to photons with a frequency $\omega ^{\prime
} $ -- are constrained by the values of $\alpha .$ Then, integrating on
constrained values of $\alpha $, i.e. $\alpha \in \left[ 0,\alpha _{\max }%
\right] ,$ is equivalent to integrate on values $\omega $ allowed by the
initial source. Then (\ref{Jb}) writes:

\begin{equation}
J=\int_{V_{\omega ^{\prime }}}\omega ^{\prime }\left( \frac{d\sigma }{%
dS^{\prime }}\right) \times \Phi \left( \omega \right) \left( \frac{d\omega 
}{d\omega ^{\prime }}\right) \times n_{\varphi }r^{2}dr\left( \frac{d\Omega 
}{d\alpha }\right) d\alpha ,  \label{J}
\end{equation}%
which can be written as:%
\begin{eqnarray}
J &=&\frac{5\alpha _{\text{em}}^{4}}{72\pi m^{4}}\omega ^{\prime
3}K_{\varphi }\int_{0}^{\alpha _{\max }}d\alpha   \label{Jexp} \\
&&\frac{\left( 1-\cos \alpha \right) ^{2}\sin \alpha }{\left( \left( E-\sqrt{%
E^{2}-m_{\varphi }^{2}}\cos \theta \right) -\omega ^{\prime }\left( 1-\cos
\alpha \right) \right) \cos \alpha }  \notag \\
&&\times B\left( \frac{\omega ^{\prime }\left( E-\sqrt{E^{2}-m_{\varphi }^{2}%
}\cos \left( \alpha -\theta \right) \right) }{\left( E-\sqrt{%
E^{2}-m_{\varphi }^{2}}\cos \theta \right) -\omega ^{\prime }\left( 1-\cos
\alpha \right) }\right) ,  \notag
\end{eqnarray}%
by using Eqs. (\ref{ds}) and (\ref{wp}), and with:

\begin{equation}
\cos \alpha _{\max }=1-\frac{E-\sqrt{E^{2}-m_{\varphi }^{2}}\cos \theta }{%
\omega ^{\prime }},  \label{max}
\end{equation}%
and 
\begin{equation}
K_{\varphi }=\int_{R_{0}}^{\infty }\left( \frac{R_{0}}{r}\right)
^{2}n_{\varphi }dr,  \label{Kphi}
\end{equation}%
where:%
\begin{equation}
K_{\varphi }\sim 0.32\,n_{\varphi ,local}R_{0},  \label{k2}
\end{equation}%
for the NFW density profile.

In the limit $M_{B}<10^{4}$ TeV, (\ref{Jexp}) writes: 
\begin{equation}
J=J_{0}F\left( \overline{\omega }\right) ,  \label{Jf}
\end{equation}%
with $\overline{\omega }=\omega ^{\prime }/T$, and where:

\begin{equation}
J_{0}=\frac{5\alpha _{\text{em}}^{4}}{144\pi ^{4}m^{4}}K_{\varphi
}E^{2}T^{3},  \label{J0}
\end{equation}%
and:%
\begin{equation}
F\left( \overline{\omega }\right) =\int_{0}^{1}\frac{y^{5}}{\left(
1-y^{2}\right) ^{4}}\frac{\overline{\omega }^{3}}{\exp \left( \overline{%
\omega }\frac{1}{1-y^{2}}\right) -1}dy.  \label{F}
\end{equation}%
The behavior of $F\left( \overline{\omega }\right) $ is roughly a decaying
exponential, and it can be shown that:%
\begin{equation}
F\left( \overline{\omega }\right) \sim 1.2e^{-\overline{\omega }},
\label{approx}
\end{equation}%
with a relative error lower than $20$ \%.

\subsection*{Conditions and possibility of observation}

Let us now consider a typical giant blue star. This choice is motivated by a huge star temperature and a large star diameter, allowing for a large value of $J_{0}$. Let us consider for instance Deneb, such that $R_{0}\sim200 R_{\odot}$ and $T=8700$ K \cite{deneb}. One uses $n_{\varphi,local}\sim2.2\times10^{13}$ m$^{-3}$ and $E\sim m_{\varphi}\approx35$ meV as defined in the previous sections, as a set of upper values reachable in our model. Then:
\begin{equation}
    J_{0}\sim1.3\times10^{-33}W~m^{-2}Hz^{-1}=1.3\times10^{-7} \text{ Jy.} \label{eq:J0_value}
\end{equation}
Despite its low magnitude, this value falls within the sensitivity range of current state-of-the-art instruments \cite{Kalas2013,Guyon2018,Hinkley2022,ins1,ins2,ins3,Kalas2008,Bowler2016}. The NFW profile's concentration near the star enhances the signal at small angular separations, ideal for coronagraphic observations \cite{Guyon2018,Bowler2016,ins3} with instruments like the James Webb Space Telescope (JWST) or the upcoming Extremely Large Telescope (ELT) \cite{Kalas2013,Guyon2018,Hinkley2022,ins1,ins2}.\\

The primary challenge in detecting such a faint halo lies in distinguishing its signal from astrophysical backgrounds and instrumental noise. The dominant astrophysical backgrounds are expected to be diffuse light from circumstellar dust scattering and line emission from circumstellar gas \cite{Kalas2008}. For a star like Deneb, these sources can produce fluxes in a comparable range, from $10^{-8}$ to $10^{-6}$ Jy \cite{Kalas2013}.

However, the scalar boson signal possesses a key distinguishing feature: its unique spectral shape. Whereas dust scattering typically produces a modified blackbody spectrum or a power-law dependence on frequency (e.g., $\propto \omega^2$ in the Rayleigh regime), and gas emission is confined to sharp, narrow spectral lines, the boson-scattered halo is predicted to exhibit a broad, featureless spectrum with a distinct exponential decay, $F(\overline{\omega})\sim1.2e^{-\overline{\omega}}$. This unique spectral signature provides a powerful tool for discriminating the signal from backgrounds through high-resolution spectroscopic analysis.

Furthermore, ancillary measurements could aid in this discrimination. For instance, polarimetry could help differentiate the Compton-like scattering signature of the boson from that of dust, which often produces highly polarized light \cite{Bowler2016}. Multi-wavelength observations would further constrain the nature of any detected halo, making it possible to rule out conventional astrophysical sources \cite{Hinkley2022}. The angular resolution required ($\theta \sim 0.001 \, \text{arcsec}$) is already within reach of JWST's coronagraphs in the near-infrared \cite{Hinkley2022}. While recent observations have not reported such a halo \cite{Hinkley2022}, targeted spectroscopic analysis of existing or future datasets could place the first direct upper limits on $J_0$ and, consequently, on the parameters of the baryogenesis model \cite{baryo}.

\section{Conclusion}

\label{conclusion}

In this work, one investigated the phenomenological viability of 
detecting a pseudo-scalar boson predicted by a recent braneworld-based 
baryogenesis scenario \cite{baryo}. One has shown that these bosons, 
produced in the early Universe, would persist today as a subdominant 
component of dark matter. While their contribution to the cosmic density 
is negligible, one demonstrated that a one-loop quantum process allows 
for a photon-boson interaction, leading to a potentially observable signature.

The main result of our study is the prediction that light from massive
blue stars, when scattered by these relic bosons, would generate a faint 
halo with a distinctive, exponentially decaying spectral signature. The detection 
and analysis of this specific spectrum, or the placing of upper limits 
on its intensity, would not serve to identify a major dark matter candidate, 
but rather to provide a direct and powerful experimental constraint on the 
properties of the scalar field and, by extension, on the braneworld baryogenesis 
model itself \cite{baryo}. This approach opens a new, observationally-driven 
window to probe the physics of extra dimensions and the origin of matter 
in our Universe. Future observations with next-generation telescopes could 
further test the viability of this scenario.

\begin{appendix}

\section{Photon--pseudo-scalar boson scattering amplitude}

\label{appendixA}

Following the Lagrangian (\ref{Lag}), there is no direct coupling between
the electromagnetic field and the scalar boson. Nevertheless, it is possible
to build a photon-boson coupling at the loop level as shown in Fig. 3. Here
one computes the amplitude and the average squared amplitude related to such
a process. As shown in Fig. 3, only four box diagrams are allowed. The
amplitude $\mathcal{M}$ can be expressed as a sum of four integrals as:

\begin{equation}
\mathcal{M}=\frac{-i}{4}e^{4}\varepsilon _{\nu }^{\ast }(p_{3},\lambda
_{2})\varepsilon _{\mu }(p_{1},\lambda _{1})\mathcal{F}^{\nu \mu
}(p_{1},p_{2},p_{3},p_{4}),  \label{A1}
\end{equation}%
with:%
\begin{equation}
\mathcal{F}^{\nu \mu }=\mathcal{F}_{sq}^{\nu \mu }+\mathcal{F}_{\overline{sq}%
}^{\nu \mu }+\mathcal{F}_{cr}^{\nu \mu }+\mathcal{F}_{\overline{cr}}^{\nu
\mu},  \label{A2}
\end{equation}%
where $sq$ (respectively $cr$) is for the square diagrams (respectively the
crossed diagrams). The bar over $sq$ (respectively $cr$) is for diagrams
with reversed momenta. One gets:

\begin{widetext}
\begin{eqnarray}
\mathcal{F}_{sq}^{\nu \mu } &=&\frac{1}{\left( 2\pi \right) ^{4}}\int d^{4}k%
\frac{\text{Tr}\left[ \left( -\slashed{k}+\slashed{p}_{2}+m\right) \gamma
^{5}\left( -\slashed{k}+m\right) \gamma ^{5}\left( \slashed{p}_{1}+%
\slashed{p}_{2}-\slashed{p}_{3}-\slashed{k}+m\right) \gamma ^{\nu }\left( %
\slashed{p}_{1}+\slashed{p}_{2}-\slashed{k}+m\right) \gamma ^{\mu }\right] }{%
\left( \left( -k+p_{2}\right) ^{2}-m^{2}\right) \left( k^{2}-m^{2}\right)
\left( \left( p_{1}+p_{2}-p_{3}-k\right) ^{2}-m^{2}\right) \left( \left(
p_{1}+p_{2}-k\right) ^{2}-m^{2}\right)},  \label{A2a} \\
\mathcal{F}_{\overline{sq}}^{\nu \mu } &=&\frac{1}{\left( 2\pi \right) ^{4}}%
\int d^{4}k\frac{\text{Tr}\left[ \left( \slashed{k}-\slashed{p}_{2}+m\right)
\gamma ^{5}\left( \slashed{k}+m\right) \gamma ^{5}\left( -\slashed{p}_{1}-%
\slashed{p}_{2}+\slashed{p}_{3}+\slashed{k}+m\right) \gamma ^{\nu }\left( -%
\slashed{p}_{1}-\slashed{p}_{2}+\slashed{k}+m\right) \gamma ^{\mu }\right] }{%
\left( \left( -k+p_{2}\right) ^{2}-m^{2}\right) \left( k^{2}-m^{2}\right)
\left( \left( p_{1}+p_{2}-p_{3}-k\right) ^{2}-m^{2}\right) \left( \left(
p_{1}+p_{2}+k\right) ^{2}-m^{2}\right)},  \label{A2b} \\
\mathcal{F}_{cr}^{\nu \mu } &=&\frac{1}{\left( 2\pi \right) ^{4}}\int d^{4}k%
\frac{\text{Tr}\left[ \left( \slashed{p}_{2}-\slashed{p}_{3}-\slashed{k}%
+m\right) \gamma ^{\nu }\left( -\slashed{k}+\slashed{p}_{2}+m\right) \gamma
^{5}\left( -\slashed{k}+m\right) \gamma ^{5}\left( \slashed{p}_{1}+%
\slashed{p}_{2}-\slashed{p}_{3}-\slashed{k}+m\right) \gamma ^{\mu }\right] }{%
\left( \left( p_{2}-p_{3}-k\right) ^{2}-m^{2}\right) \left( \left(
-k+p_{2}\right) ^{2}-m^{2}\right) \left( k^{2}-m^{2}\right) \left( \left(
p_{1}+p_{2}-p_{3}-k\right) ^{2}-m^{2}\right)},  \label{A2c} \\
\mathcal{F}_{\overline{cr}}^{\nu \mu } &=&\frac{1}{\left( 2\pi \right) ^{4}}%
\int d^{4}k\frac{\text{Tr}\left[ \left( -\slashed{p}_{2}+\slashed{p}_{3}+%
\slashed{k}+m\right) \gamma ^{\nu }\left( \slashed{k}-\slashed{p}%
_{2}+m\right) \gamma ^{5}\left( \slashed{k}+m\right) \gamma ^{5}\left( -%
\slashed{p}_{1}-\slashed{p}_{2}+\slashed{p}_{3}+\slashed{k}+m\right) \gamma
^{\mu }\right] }{\left( \left( p_{2}-p_{3}-k\right) ^{2}-m^{2}\right) \left(
\left( -k+p_{2}\right) ^{2}-m^{2}\right) \left( k^{2}-m^{2}\right) \left(
\left( p_{1}+p_{2}-p_{3}-k\right) ^{2}-m^{2}\right)}.  \label{A2d}
\end{eqnarray}%
\end{widetext}

Since the incident scalar boson is expected to get a very low momentum in
comparison with the photons of interest, the expression of the amplitude can
be simplified through a soft approximation, i.e. $p_{2}=0$ inside the loop.
Then, using commutation properties of $\gamma ^{5}$ with $\gamma ^{\mu }$
matrices, and the fact that the trace is invariant under cyclic permutation
of the operators, one obtains:

\begin{widetext}
\begin{eqnarray}
\mathcal{F}_{sq}^{\nu \mu } &=&\frac{-1}{\left( 2\pi \right) ^{D}}\mu
^{4-D}\int d^{D}k\frac{\text{Tr}\left[ \left( \slashed{p}_{1}-\slashed{p}%
_{3}-\slashed{k}+m\right) \gamma ^{\nu }\left( \slashed{p}_{1}-\slashed{k}%
+m\right) \gamma ^{\mu }\right] }{\left( k^{2}-m^{2}\right) \left( \left(
p_{1}-p_{3}-k\right) ^{2}-m^{2}\right) \left( \left( p_{1}-k\right)
^{2}-m^{2}\right)},  \label{A3a} \\
\mathcal{F}_{\overline{sq}}^{\nu \mu } &=&\frac{-1}{\left( 2\pi \right) ^{D}}%
\mu ^{4-D}\int d^{D}k\frac{\text{Tr}\left[ \left( \slashed{p}_{1}-\slashed{p}%
_{3}-\slashed{k}-m\right) \gamma ^{\nu }\left( \slashed{p}_{1}-\slashed{k}%
-m\right) \gamma ^{\mu }\right] }{\left( k^{2}-m^{2}\right) \left( \left(
p_{1}-p_{3}-k\right) ^{2}-m^{2}\right) \left( \left( p_{1}-k\right)
^{2}-m^{2}\right)},  \label{A3b} \\
\mathcal{F}_{cr}^{\nu \mu } &=&\frac{-1}{\left( 2\pi \right) ^{D}}\mu
^{4-D}\int d^{D}k\frac{\text{Tr}\left[ \left( \slashed{p}_{1}-\slashed{p}%
_{3}-\slashed{k}+m\right) \gamma ^{\mu }\left( -\slashed{p}_{3}-\slashed{k}%
+m\right) \gamma ^{\nu }\right] }{\left( k^{2}-m^{2}\right) \left( \left(
p_{1}-p_{3}-k\right) ^{2}-m^{2}\right) \left( \left( -p_{3}-k\right)
^{2}-m^{2}\right)},  \label{A3c} \\
\mathcal{F}_{\overline{cr}}^{\nu \mu } &=&\frac{-1}{\left( 2\pi \right) ^{D}}%
\mu ^{4-D}\int d^{D}k\frac{\text{Tr}\left[ \left( \slashed{p}_{1}-\slashed{p}%
_{3}-\slashed{k}-m\right) \gamma ^{\mu }\left( -\slashed{p}_{3}-\slashed{k}%
-m\right) \gamma ^{\nu }\right] }{\left( k^{2}-m^{2}\right) \left( \left(
p_{1}-p_{3}-k\right) ^{2}-m^{2}\right) \left( \left( -p_{3}-k\right)
^{2}-m^{2}\right)},  \label{A3d}
\end{eqnarray}%
\end{widetext}
and where one has introduced a dimensional regularization, with $\mu $ a
mass parameter introduced to ensure the dimensionality of the amplitude.

It is now relevant to consider the expression of $\mathcal{F}^{\nu \mu }$
when all the legs momenta vanish. One easily gets:%
\begin{eqnarray}
&&\mathcal{F}^{\nu \mu }(0,0,0,0)  \label{A4} \\
&=&\frac{-16}{\left( 2\pi \right) ^{D}}\mu ^{4-D}\int d^{D}k\frac{\left(
-k^{2}+m^{2}\right) g^{\mu \nu }+2k^{\nu }k^{\mu }}{\left(
k^{2}-m^{2}\right) ^{3}}.  \notag
\end{eqnarray}%
Then, using the well-known one-loop master formula \cite{Book1}:%
\begin{eqnarray}
&&\frac{1}{\left( 2\pi \right) ^{D}}\int d^{D}k\frac{\left( -k^{2}\right)
^{a}}{\left( -k^{2}+m^{2}\right) ^{b}}  \label{A5} \\
&=&\frac{i\pi ^{D/2}}{\left( 2\pi \right) ^{D}}\frac{\Gamma (\frac{D}{2}%
+a)\Gamma (b-a-\frac{D}{2})}{\Gamma (\frac{D}{2})\Gamma (b)}\frac{1}{%
m^{2\left( b-a-\frac{D}{2}\right)}},  \notag
\end{eqnarray}%
and%
\begin{eqnarray}
&&\frac{1}{\left( 2\pi \right) ^{D}}\int d^{D}k\frac{-k^{\nu }k^{\mu }}{%
\left( -k^{2}+m^{2}\right) ^{b}}  \label{A6} \\
&=&\frac{i\pi ^{D/2}}{\left( 2\pi \right) ^{D}}\frac{\Gamma (b-1-\frac{D}{2})%
}{2\Gamma (b)}\frac{g^{\mu \nu }}{m^{2\left( b-1-\frac{D}{2}\right)}}. 
\notag
\end{eqnarray}%
Eqs. (\ref{A4}) then becomes:%
\begin{eqnarray}
&&\mathcal{F}^{\nu \mu }(0,0,0,0)  \label{A7} \\
&=&8g^{\mu \nu }\mu ^{4-D}\frac{i\pi ^{D/2}}{\left( 2\pi \right) ^{D}}\frac{1%
}{m^{2\left( 2-\frac{D}{2}\right) }}\Gamma \left( 2-\frac{D}{2}\right), 
\notag
\end{eqnarray}%
such that:%
\begin{eqnarray}
&&\lim_{D\rightarrow 4}\mathcal{F}^{\nu \mu }(0,0,0,0)  \label{A8} \\
&=&g^{\mu \nu }\frac{i}{2\pi ^{2}}\left( -\frac{2}{D-4}-\gamma _{E}\right) +%
\mathcal{O}(D-4).  \notag
\end{eqnarray}%
Thus, $\mathcal{F}^{\nu \mu }(0,0,0,0)$, but also $\mathcal{F}^{\nu \mu }$,
gets a pole for $D=4$. Then, the amplitude can be renormalized by
subtracting $\mathcal{F}^{\nu \mu }(0,0,0,0)$ from $\mathcal{F}^{\nu \mu }$%
, following the usual 't Hooft-Veltman procedure \cite{Book2, thooft1}, thus
allowing that the amplitude vanishes when the four-momenta of the legs tend
towards zero. Then, after renormalization, one easily gets:

\begin{widetext}
\begin{eqnarray}
\mathcal{F}_{sq}^{\nu \mu }+\mathcal{F}_{\overline{sq}}^{\nu \mu } &=&\frac{%
-2}{\left( 2\pi \right) ^{D}}\mu ^{4-D}\int d^{D}k\frac{\text{Tr}\left[
\left( \slashed{p}_{1}-\slashed{p}_{3}\right) \gamma ^{\nu }\left( %
\slashed{p}_{1}-\slashed{k}\right) \gamma ^{\mu }-\slashed{k}\gamma ^{\nu }%
\slashed{p}_{1}\gamma ^{\mu }\right] }{\left( k^{2}-m^{2}\right) \left(
\left( p_{1}-p_{3}-k\right) ^{2}-m^{2}\right) \left( \left( p_{1}-k\right)
^{2}-m^{2}\right)},  \label{A9a} \\
\mathcal{F}_{cr}^{\nu \mu }+\mathcal{F}_{\overline{cr}}^{\nu \mu } &=&\frac{%
-2}{\left( 2\pi \right) ^{D}}\mu ^{4-D}\int d^{D}k\frac{\text{Tr}\left[
\left( \slashed{p}_{1}-\slashed{p}_{3}\right) \gamma ^{\mu }\left( -%
\slashed{p}_{3}-\slashed{k}\right) \gamma ^{\nu }+\slashed{k}\gamma ^{\mu }%
\slashed{p}_{3}\gamma ^{\nu }\right] }{\left( k^{2}-m^{2}\right) \left(
\left( p_{1}-p_{3}-k\right) ^{2}-m^{2}\right) \left( \left( -p_{3}-k\right)
^{2}-m^{2}\right)}.  \label{A9b}
\end{eqnarray}
\end{widetext}

Using a Veltman-Passarino-like reduction \cite{pv}, one sets: 
\begin{equation}
K_{\mu }(p_{1})=p_{1\mu }C_{0}-S_{\mu },  \label{A10a}
\end{equation}%
and%
\begin{equation}
K_{\mu }(-p_{3})=-p_{3\mu }C_{0}+S_{\mu },  \label{A10b}
\end{equation}%
with

\begin{widetext}
\begin{equation}
C_{0}=\frac{-2\mu ^{4-D}}{\left( 2\pi \right) ^{D}}\int d^{D}k\frac{1}{%
\left( k^{2}-m^{2}\right) \left( \left( p_{1}-k\right) ^{2}-m^{2}\right)
\left( \left( p_{3}-k\right) ^{2}-m^{2}\right) },  \label{A11}
\end{equation}%
and%
\begin{equation}
S_{\mu }=\frac{-2\mu ^{4-D}}{\left( 2\pi \right) ^{D}}\int d^{D}k\frac{%
k_{\mu }}{\left( k^{2}-m^{2}\right) \left( \left( p_{1}-k\right)
^{2}-m^{2}\right) \left( \left( p_{3}-k\right) ^{2}-m^{2}\right)}.
\label{A12}
\end{equation}%
Eqs. (\ref{A9a}) and (\ref{A9b}) then immediately writes as: 
\begin{eqnarray}
\mathcal{F}_{sq}^{\nu \mu }+\mathcal{F}_{\overline{sq}}^{\nu \mu } &=&\text{%
Tr}\left[ \left( \slashed{p}_{1}-\slashed{p}_{3}\right) \gamma ^{\nu }%
\slashed{S}\gamma ^{\mu }-\slashed{K}(p_{1})
\gamma ^{\nu }\slashed{p}_{1}\gamma ^{\mu }\right],  \label{A13a} \\
\mathcal{F}_{cr}^{\nu \mu }+\mathcal{F}_{\overline{cr}}^{\nu \mu } &=&-\text{%
Tr}\left[ \left( \slashed{p}_{1}-\slashed{p}_{3}\right) \gamma ^{\mu }%
\slashed{S}\gamma ^{\nu }-\slashed{K}(-p_{3})
\gamma ^{\mu }\slashed{p}_{3}\gamma ^{\nu }\right],  \label{A13b}
\end{eqnarray}%
from which one deduces after trace calculation:%
\begin{equation}
\mathcal{F}^{\nu \mu }=-8C_{0}\left( p_{1}^{\nu }p_{1}^{\mu }+p_{3}^{\nu
}p_{3}^{\mu }\right) +4\left( -g^{\mu \nu }(p_{1}+p_{3})S+(p_{1}+p_{3})^{\nu
}S^{\mu }+(p_{1}+p_{3})^{\mu }S^{\nu }\right).  \label{A14}
\end{equation}%
\end{widetext}

It can be easily check that the Ward identities $\mathcal{F}^{\nu \mu
}p_{1\mu }=\mathcal{F}^{\nu \mu }p_{3\nu }=0$ are strictly verified if $%
p_{3}=p_{1}$ only. Here, they are asymptotically verified in the soft
approximation as $p_{3}\approx p_{1}$, since $p_{2}\ll p_{1},$ $p_{3}$. That
means that the one-loop approximation here considered will not be valid for
photon energies close to those of the scalar boson. 
Anyway, given the anticipated low values for $m_\phi$ and $E$ (see section \ref{Relic}) 
relative to the expected photon energies (see section \ref{star}), this condition is not 
particularly restrictive. Therefore, it is considered to be consistently satisfied 
within the scope of this study. Otherwise, Wilson lines would need to be introduced 
in loop diagrams but at the cost of a substantial complication far beyond the scope of the present paper.

Using Eq. (\ref{A1}), the average squared 
amplitude $\overline{\left\vert \mathcal{M}\right\vert ^{2}}$ over the 
initial photon polarization, and summed over the final photon polarization, writes as:
\begin{equation}
\overline{\left\vert \mathcal{M}\right\vert ^{2}}=\frac{1}{2}%
\sum_{polarizations}\left\vert \mathcal{M}\right\vert ^{2}=\frac{1}{32}e^{8}%
\mathcal{F}_{\nu \mu }^{\ast }\mathcal{F}^{\nu \mu},  \label{A15}
\end{equation}%
where one used the usual substitution $\sum_{polarizations}\varepsilon _{\nu
}^{\ast }\varepsilon _{\mu }\rightarrow -g_{\nu \mu }$ \cite{Book2}. From
Eq. (\ref{A14}) one gets (for $D=4$): 
\begin{widetext}
\begin{eqnarray}
\mathcal{F}_{\nu \mu }^{\ast }\mathcal{F}^{\nu \mu } &=&128\left\vert
C_{0}\right\vert ^{2}\left( p_{1}p_{3}\right) ^{2}+32\left\vert
(p_{1}+p_{3})S\right\vert ^{2}+64\left( p_{1}p_{3}\right) S_{\mu }^{\ast
}S^{\mu }  \label{A16} \\
&&-64C_{0}^{\ast }\left( p_{1}p_{3}\right) \left( \left( p_{1}+p_{3}\right)
S\right) -64C_{0}\left( p_{1}p_{3}\right) \left( \left( p_{1}+p_{3}\right)
S^{\ast }\right),  \notag
\end{eqnarray}
\end{widetext}
It is then relevant to estimate Eq. (\ref{A16}) for small values of $p_{1},$ 
$p_{3}$ (compared with $m$). From one-loop master expression (\ref{A5}), one notes that: 
\begin{equation}
\lim_{p_{1},p_{3}\rightarrow 0\text{ ; }D\rightarrow 4}C_{0}=\frac{i}{16\pi
^{2}}\frac{1}{m^{2}}.  \label{A17}
\end{equation}%
Also, Eq. (\ref{A12}) can be easily described by the relevant first order
taylor approximation $S_{\mu }\sim S_{\mu }(0)+\left( \frac{\partial S_{\mu }%
}{\partial p_{1\nu }}\right) _{0}p_{1\nu }+\left( \frac{\partial S_{\mu }}{%
\partial p_{3\nu }}\right) _{0}p_{3\nu }$, such that:

\begin{equation}
S\sim \frac{i}{48\pi ^{2}}\frac{1}{m^{2}}\left( p_{1}+p_{3}\right).
\label{A18}
\end{equation}%
Then, using Eqs. (\ref{A15}), (\ref{A16}), (\ref{A17}) and (\ref{A18}), the
average squared amplitude finally writes:%
\begin{equation}
\overline{\left\vert \mathcal{M}\right\vert ^{2}}=\frac{20}{9}\frac{\alpha _{%
\text{em}}^{4}}{m^{4}}(p_{1}p_{3})^{2}, \label{A19}
\end{equation}%
where $\alpha _{\text{em}}=e^{2}/(4\pi )$ is the fine-structure constant.

\end{appendix}

\end{document}